\documentstyle[prd,aps,multicol,tighten,epsf,rotate]{revtex}

\begin{document}
\draft

\preprint{\vbox{\it 
                        \null\hfill\rm    IP-BBSR/97-19, April'97}\\\\}
%
\title{\Large\bf Vortex-Antivortex Pair Production in a 
First Order Phase Transition}
\author{\large Sanatan Digal, Supratim Sengupta and Ajit M. 
Srivastava \footnote{E-mail :\\
digal@iopb.ernet.in \\ supratim@iopb.ernet.in \\ ajit@iopb.ernet.in}}
\address{\large\it Institute of Physics\\
Sachivalaya Marg, Bhubaneswar--751005, INDIA}
\maketitle
\widetext
\parshape=1 0.75in 5.5in
\begin{abstract}
We carry out numerical simulation of a first order phase transition in 
2+1 dimensions by randomly nucleating bubbles, and study the formation 
of global U(1) vortices. Bubbles grow and coalesce and vortices 
are formed at junctions of bubbles via standard Kibble mechanism 
as well as due to a new mechanism, recently proposed by us, where 
defect-antidefect pairs are produced due to field oscillations.  
We make a comparative study of the contribution of both of
these mechanisms for vortex production. We find that, for high
nucleation rate of bubbles, vortex-antivortex pairs produced via 
the new mechanism have overlapping configurations, and annihilate quickly;
so only those vortices survive till late which are produced via the 
Kibble mechanism. However, for low nucleation rates, bubble collisions 
are energetic enough to lead to many well separated 
vortex-antivortex pairs being produced via the new mechanism. For 
example, in a simulation involving nucleation of 20 bubbles, a total of 
14 non-overlapping vortices and antivortices formed via this new mechanism 
of pair creation (6 of them being very well separated), as compared to
6 vortices and antivortices produced via the Kibble 
mechanism. Our results show the possibility that in extremely energetic 
bubble collisions, such as those in the inflationary models of the 
early Universe, this new mechanism may drastically affect the defect 
production scenario. 
\end{abstract}
\vskip 0.125 in
\parshape=1 -.75in 5.5in
\begin{multicols}{2}
\narrowtext

\vskip .2in
\centerline {\bf 1. Introduction}
\vskip .1in

 Production of topological defects has been a subject of
great interest to condensed matter physicists, as well as to
particle physicists in the context of the models of the early 
Universe \cite{shlrd}. Conventionally there have been two 
types of processes thought to be responsible for the production 
of defects. Pair production of defects-antidefects via thermal 
fluctuations \cite{thrm} is one of them while the second 
process, usually known as the Kibble mechanism \cite{kbl}, arises
from the formation of a domain structure after the phase
transition. In the context of the early Universe, it is
the Kibble mechanism which plays the dominant role as
thermally produced defects are generally Boltzmann suppressed.
 
 Recently, we have proposed a new mechanism for defect production
which arises due to field oscillations \cite{dss}. This mechanism
was first discussed by two of us in the context of systems
with small explicit symmetry breaking terms \cite{dgl}, for the 
case when transition is of first order.  There are many examples 
of such systems, such as axionic strings and Skyrmions for particle 
physics and liquid crystal defects in the presence of external 
fields for condensed matter systems. Subsequently, we showed 
that this mechanism is not limited to systems with explicit 
symmetry breaking. We analyzed the underlying physics
of the mechanism and showed that this mechanism is completely
general \cite{dss}. It applies to the production of all sorts of
topological defects and even for second order transitions
involving quench from high temperatures.

 With the demonstration of the general applicability of this 
mechanism, it becomes important to ask about its relative 
importance in determining the defect distribution arising
in a phase transition. The numerical simulations (for the 
first order transition case) carried out in \cite{dss,dgl} 
considered certain {\it specific} field configurations of
bubbles as the initial conditions and showed, in 
detail, how this new mechanism actually operates, and how
much enhancement in vortex production may occur in certain
favorable conditions. For example, for the explicit symmetry
breaking case in \cite{dgl} it was shown that in certain
cases one may get up to ten vortices and antivortices produced 
from a single two bubble collision. Similarly, in \cite{dss} (for
the case when no explicit symmetry breaking is present),
it was shown that with certain favorable distribution of
phases and bubble separation, vortex-antivortex pair may form 
via this mechanism which is as well separated as the ones which
are typically produced via the Kibble mechanism.

  However, as these initial conditions were specially chosen,
it only shows the possibility that this mechanism {\it may}
play an important role in phase transitions. What one would
like to know is the actual contribution of this
mechanism for defect production in a phase transition where
bubbles are randomly nucleated, because that is the only
quantity which is of experimental interest. We address this
problem in this paper and focus on the case of most general
interest, when there is no explicit symmetry breaking involved.
The case with the explicit symmetry breaking is very
specialized and the dynamics of vortex production also
very different from the case with no explicit symmetry
breaking; even though the underlying mechanism is still the 
same, arising due to oscillations of the field.  Explicit symmetry 
breaking leads to extra features in the dynamics which play a very
crucial role in determining the defect abundance. For example, in 
presence of explicit symmetry breaking, the energetics of field 
oscillations in the 
coalesced portion of the bubbles is governed not only by the 
energy acquired by the bubble walls due to conversion of false 
vacuum to true vacuum, but also due to the energy stored in 
the bubble walls due to the tilt in effective potential.
In order that such effects do not obscure the main point we 
are trying to study, which is to know the relative importance 
of this new mechanism in a general phase transition, we will 
only focus, in this paper, on systems with spontaneously broken
global symmetry {\it without} any explicit symmetry breaking. 
In any case, it is these systems which form the most general class
of systems where defect production is of interest, especially 
in the context of structure formation in the early Universe. We will 
present the study of defect production in phase transitions for the 
case of explicit symmetry breaking in a followup work \cite{dss2}. 

 The paper is organized in the following manner. The second section presents 
the essential physical picture of this mechanism by reviewing earlier results 
from \cite{dss}. Section 3 discusses the numerical technique used for 
implementing the phase transition by random nucleation of bubbles. We discuss 
the algorithm for detecting vortices produced in the transition. Each vortex
is later analyzed in detail to make sure of its structure  as well as the 
specific mechanism responsible for its production. We present this analysis 
and other numerical results in Section 4, where we identify vortices 
which are produced via the Kibble mechanism and those which are produced 
via the new mechanism. Section 5 provides discussion of the results where 
we compare vortex production via these two mechanisms and discuss various 
implications of the new mechanism. We argue that, due to this mechanism, in 
first order transitions with low nucleation rates (as would be the case for 
inflationary scenarios of the early Universe) violent bubble collisions 
may dramatically alter the production of defects. Conclusions are
presented in Section 5 where we summarize the main features of our results. 

\vskip .3in
\centerline{\bf 2. Physical Picture of the Mechanism}
\vskip .1in

   As we mentioned above, the production of defect-antidefect
pairs via this mechanism happens entirely due to the oscillations
of the magnitude of the order parameter field. We briefly review the
essential physical picture underlying this mechanism, see \cite{dss}.
We will study the formation of U(1) global vortices in 2+1 dimensions, 
with the order parameter being the vacuum expectation value  of
a complex scalar field $\Phi$. Consider a region of space in which the 
phase $\theta$ of the order parameter field varies uniformly from 
$\alpha$ to $\beta$ as shown in Fig.1a. At this stage there is no 
vortex or anti-vortex present in this region. Now suppose that the
magnitude of the order parameter field undergoes oscillations,
resulting in the passage of $\Phi$ through zero, in a small region in 
the center enclosed by the dotted loop, see Fig.1b. [As discussed in
\cite{dss,dgl}, and as we will see later in this paper, such 
oscillations can easily result during coalescence of bubbles in a
first order phase transition. It can also happen in second order
transitions involving quenching from high temperatures.] From a
plot of the effective potential like that of a Mexican hat, it
is easy to see that oscillation of the order parameter through 
zero magnitude amounts to a change in the order parameter
to the diametrically opposite point on the vacuum manifold $S^1$.
This process, which causes a discontinuous change in $\theta$ by 
$\pi$, was termed as the flipping of $\Phi$ in \cite{dss}. 

  For simplicity, we take $\theta$ to be uniform in the flipped region. 
Consider now the variation of $\theta$ along the closed path AOBCD (shown 
by the solid curve in Fig.1b) and assume that  $\theta$ varies along the
shortest path on the vacuum manifold $S^1$ (as indicated by the dotted 
arrows), as we cross the dotted curve i.e. the variation of $\theta$ 
from the unflipped to the flipped region follows the geodesic rule. [Even
if $\theta$ varies along the longer path on $S^1$, we still get a pair,
with the locations of the vortex and the antivortex getting interchanged.]
It is then easy to see that $\theta$ winds by 2$\pi$ as we go around the 
closed path, showing that a vortex has formed inside the region. As the 
net winding surrounding the flipped region is zero, it follows that an 
anti-vortex has formed in the other half of the dotted region. One can also 
see it by explicitly checking for the (anti)winding of $\theta$.  

 Another way to see how flipping of $\Phi$ results in the formation of a  
vortex-antivortex pair is as follows. Consider the variation of $\theta$
around the closed path AOBCD in Fig.1b before flipping of $\Phi$ in the
dotted region. Such a variation of $\theta$ corresponds to a shrinkable
loop on the vacuum manifold $S^{1}$. After flipping of $\Phi$ in the dotted
region, a portion in the center of the arc P connecting $\theta$=$\alpha$
to $\theta$=$\beta$ on $S^{1}$ moves to the opposite side of $S^{1}$.
If the midpoint of the arc originally corresponded to $\theta$=$\gamma$,
flipping of $\Phi$ changes $\gamma$ to $\gamma$+$\pi$. We assume that
different points on the arc move to the opposite side of $S^1$ maintaining
symmetry about the mid point of the arc, (and say, also maintaining
the orientation of the arc). Then one can see that the loop on the
vacuum manifold $S^1$ becomes non-shrinkable, and has winding number
one, see \cite{dss} for details.  Thus a vortex has formed inside the 
region enclosed by the solid curve. Obviously, an anti-vortex will form 
in the left half of Fig.1b. 

We should mention here that every successive passage of $\Phi$ through zero
will create a new vortex-antivortex pair. Density waves generated during
field oscillations lead to further separation of a vortex-antivortex pair 
created earlier. The attractive force between the vortex and anti-vortex 
lead to their eventual annihilation. Though, in a rapidly expanding
early Universe, it is possible that the defect and antidefect may
keep moving apart due to expansion. We emphasize that, as
argued in \cite{dss}, this mechanism is valid even for second order phase
transitions, brought about by a quench from very high temperatures,
and is also applicable for the formation of other topological defects.  
For example, it was shown in \cite{dss} that this mechanism also applies to
the production of monopoles as well as textures. For string production
in 3+1 dimensions, above arguments can easily be seen to lead to the
production of string loops enclosing the oscillating region \cite{dss}.

 We mention here that there are vacuum manifolds for which the opposite
orientations of the order parameter field are identified; for example 
liquid crystals with the vacuum manifold being $RP^2$. In such cases, 
flipping of the order parameter field does not change its configuration,
implying that this mechanism may not be applicable there under general
situations. Though, it is possible to argue that in the presence of explicit
symmetry breaking this mechanism should still be applicable, especially
if the system is dissipative. This is because, for an order parameter 
configuration varying smoothly around the value which is energetically
most unfavourable (due to explicit symmetry breaking), the only way
to decrease the energy of the configuration is by creating a 
defect-antidefect pair as the field oscillates and passes through 
zero.  We will discuss this in more detail in \cite{dss2}.]  

\vskip .3in
\centerline{\bf 3. Numerical Techniques}
\vskip .1in

 The numerical techniques we use for bubble nucleation and time evolution
are the same as used in \cite{ajt}. In the following we provide essential
aspects of the numerical method.  We study the system described by the 
following Lagrangian in 2+1 dimensions.

\begin{equation}
{\it L} = {1 \over 2} \partial_{\mu} \Phi^{\*} \partial^{\mu} \Phi
- {1 \over 4} \phi^2 (\phi - 1)^2 +
\epsilon \phi^3 
\end{equation}

  This Lagrangian is expressed in terms of a dimensionless field $\Phi$
and appropriately scaled coordinates, with {\it $\phi$} and $\theta$ being
the magnitude and phase of the complex scalar field $\Phi$. The theory
described by this Lagrangian is that of a spontaneously broken global
U(1) symmetry. 
 
 The effective potential in Eq.(1) has a local minimum at $\phi = 0$.
The true minima occur at a non-zero value of $\phi$ and correspond to
the spontaneously broken symmetry phase. At zero temperature, the phase 
transition takes place by nucleation of bubbles of true vacuum in the 
background of false vacuum (which is at $\phi$=0) via quantum tunneling 
\cite{bbl}. Bubbles nucleate with critical size and expand, ultimately 
filling up the space. The bubble profile $\phi$ 
is obtained by solving the Euclidean field equation \cite{bbl}  

\begin{equation}
{d^2 \phi \over dr^2} + {2 \over r} {d \phi \over dr} - 
V^\prime(\phi) = 0
\end{equation}

\noindent subject to the boundary conditions $\phi(\infty)=0$ and
$d\phi/dr=0$ at $r=0$; where $V(\phi)$ is the effective potential 
in Eq.(1) and $r$ is the radial coordinate in the Euclidean space. 
In the Minkowski space, initial profile for the bubble is obtained by 
putting $t=0$ in the solution of the above equation. $\theta$ takes a 
constant value inside a given bubble. Bubble nucleation is achieved by 
replacing a region of false vacuum by the bubble profile 
(which is suitably truncated while taking care of appropriate smoothness 
of the configuration). Subsequent evolution of the initial field 
configuration is governed by the following classical field equations 
in Minkowski space 

\begin{equation}
\Box \Phi_i = - {\partial V(\Phi) \over \partial \Phi_i},~i=1,2
\end{equation}

\noindent where $\Phi = \Phi_1 + i \Phi_2$. Time derivatives of 
fields are set equal to zero at $t = 0$. 

 To simulate a full first order transition we need to nucleate
several such bubbles. This is done by randomly choosing the location
of the center of each bubble with some specified probability
per unit volume per unit time. Before nucleating a given
bubble, it is checked if the relevant region is in false vacuum
(i.e. it does not overlap with some other bubble already nucleated).
In case there is an overlap then nucleation of the new bubble
is skipped. Value of $\theta$ is randomly chosen for the interior
of each bubble.   

The simulation of the phase transition is carried out by nucleating 
bubbles on a square lattice with periodic boundary condition, i.e on a 
torus. The field configuration is evolved by using a discretized version 
of Eq.(3). Simulation is implemented by using a stabilized leapfrog 
algorithm of second order accuracy in both space and time. Physical size 
of the lattice taken is 320.0 x 320.0 with $\bigtriangleup x=0.16$i units. 
We choose $\bigtriangleup t = \bigtriangleup x/{\sqrt 2}$ which satisfies 
the Courant stability criteria. 

 Simulations were carried out on a Silicon Graphics Indigo 2 workstation 
at the Institute of Physics, Bhubaneswar.

 Bubbles are nucleated initially only, thus all the bubbles have same size 
as they expand.  During the course of the phase transition, and in the 
{\it absence} of damping, the entire energy produced as a result of the 
conversion of false vacuum to true vacuum goes to increase the kinetic 
energy of the bubble walls. As a result, the bubble walls acquire a lot of 
energy which gets dissipated when bubbles collide. In  bubble collisions 
(first studied in the context of the early Universe in\cite{hawk}) there 
are two different modes, $\phi$ oscillation mode, and $\theta$ oscillation
mode, in which the energy stored in the bubble walls can be dissipated.
The oscillations of $\phi$ (magnitude of $\Phi$) produced, when two bubbles 
collide, depend on the $\theta$ difference as well as on the separation 
between the two bubbles. If the phase difference between the two bubbles is 
large, then most of the energy stored in the bubble walls is dissipated in 
smoothening out the phase gradient in the coalesced portion of the bubble
walls and only a small amount of energy is converted to the $\phi$ 
oscillation mode. In the case of small phase difference between the two 
bubbles, a major portion of the energy of the bubble walls is converted to 
the oscillatory mode of $\phi$. If the $\phi$ oscillations are sufficiently
energetic then $\Phi$ may be able to climb the potential hill and overshoot
the value $\Phi=0$. Whenever this happens, a vortex-antivortex pair will 
be created, as we have discussed above. For a vortex-antivortex pair to 
be well formed and well separated, the value of $\phi$ should not be too 
close to zero in between the pair. This implies that $\Phi$ while passing 
through the value zero (which is the local minimum of V$(\phi)$) must be 
able to climb the potential hill in the same direction and roll down to 
the other side of V($\phi$). In the section that follows, we will be 
giving results of simulation to support this picture.
 
 The location of the vortices was determined by using an algorithm to
locate the winding number. As the phase transition nears completion
via the coalescence of bubbles, magnitude of $\Phi$ becomes non-zero
in most of the region with well defined phase $\theta$. We divide each 
plaquette in terms of two (right angle) triangles and check, for each 
such triangle, whether a non-zero winding is enclosed. A non-zero
winding is enclosed by the triangle if either of the following two
conditions are satisfied. (1) $\theta_{3}>\theta_{1}+\pi$ and 
$\theta_{3}<\theta_{2}+\pi$; for $\theta_{2}>\theta_{1}$,  or (2) 
$\theta_{3}<\theta_{1}+\pi$ and $\theta_{3}> \theta_{2}+\pi$; for 
$\theta_{2}<\theta_{1}$.  Here, $\theta_{1}$, $\theta_{2}$, and 
$\theta_{3}$ are the phases at the vertices of the triangle. 
Windings are detected only in regions where the magnitude of $\Phi$
is not too small in a small neighborhood of the triangle under
consideration. If $\Phi$ is too close to zero in a region then
that region is still mostly in the false vacuum and there is no
stability for any windings present there. After getting {\it probable} 
locations of vortices using the above algorithm, we check each of
these regions using detailed phase plots and surface plots of $\phi$
to check the winding of the vortex and to make sure that the vortex
has well defined structure. By checking similar plots at earlier as well 
as later time steps we determine whether the vortex was produced
due to oscillation, and subsequent flipping,  of $\Phi$, or via
the Kibble mechanism.

In a recent work, Copeland and Saffine have studied two bubble collisions 
for the Abelian Higgs model \cite{cs}. It is shown in \cite{cs} that
the geodesic rule in between the two bubbles is violated due to 
oscillations of $\phi$, and vortex-antivortex pair is produced in that 
region. Here the gauge fields provide a driving force for $\theta$
leading to  $\theta$ gradient in the coalesced region. More recently, 
they have also studied the formation of nontopological strings in
bubble collisions \cite{cs2}. In this context,
we would like to emphasize that the  only key ingredients for 
vortex-antivortex pair creation via the new mechanism is a region of 
varying $\theta$ with large $\phi$ oscillation in the interior. [Thus,
note that the variation of $\theta$ along a curve passing through the
flipped region, e.g. along AOB in Fig.1b, also does not follow the geodesic
rule.] Presence of other factors, such as gauge fields etc., can only affect 
the dynamical details of $\phi$ oscillations. For example, the dynamics
of $\phi$ oscillations for the case with explicit symmetry breaking 
\cite{dgl} is quite different from the case {\it without} explicit 
symmetry breaking \cite{dss}, even though the underlying mechanism
of vortex production is the same, i.e. via field oscillations.

\vskip .3in
\centerline{\bf 4. Results of the Simulation}
\vskip .1in

 In this section we describe the results of a full simulation of the phase
transition involving random nucleation of bubbles with low nucleation rate.
We have also carried out simulations with large nucleation rates, these
largely reproduce earlier results, as given in \cite{ajt}, where the
extra vortex-antivortex pairs produced were highly
overlapping and annihilated quickly. This happened because for large
nucleation rates, average separation between bubble nucleation sites is
small. Thus, bubble collisions were not energetic enough, due to low
kinetic energies of the walls, to lead to sufficiently energetic
field oscillations. In contrast, a low nucleation rate ensures that 
bubble collisions are very energetic.  This leads to an increased 
possibility of flipping of $\Phi$, thereby resulting in the creation of 
many well separated vortex-antivortex pairs, as we show below. 

 In the simulation, twenty bubbles are randomly nucleated with arbitrary 
phases chosen for bubble interiors. The bubbles expand and collide with each 
other, and vortices are formed at the junctions of three or more bubbles due 
to Kibble mechanism \cite{kbl} as well as due to flipping of $\Phi$ in 
regions where field is oscillating. In our simulations we find a total 
(time integrated) of {\it seven well separated pairs}, i.e. 14 vortices 
and antivortices, forming due to the new mechanism of flipping of $\Phi$.
In comparison, we find that 6 vortices and antivortices are produced via the
Kibble mechanism. Thus, for low nucleation rates, this new mechanism becomes 
very prominent, even for zero explicit symmetry breaking case. We count only 
those vortices which are separated by a distance which is  {\it greater} than 
the core size of the string; the core size being of the order of the inverse 
of the Higgs mass ($\simeq 2.8$ for our case). Moreover, apart from these well 
separated pairs, there were many clusters of vortex-antivortex pairs which 
were highly overlapping. We have not counted these pairs as  they annihilate 
quickly. The time for which a pair lasts depends upon many factors such as 
the presence of other vortices in the neighborhood, presence of field 
oscillations in the neighboring region which lead to density waves etc. 

 As we mentioned, we find a total of 6 vortices and antivortices which
form via the Kibble mechanism. Formation of these vortices happens
in a similar manner as was found in the simulations in \cite{ajt}.
However, the formation of vortex-antivortex pairs is now qualitatively
different. With the understanding of the precise mechanism underlying
the formation of such pairs, we are now able to focus on simulations
where formation of such pairs is the dominant process of defect creation. 
As we mentioned above, we achieve this by using a low nucleation rate for 
bubbles. Average separation between the bubble centers in the present case
is about 65 units, which is about twice of the average separation used in 
\cite{ajt}. [In the present simulation, bubble nucleation sites were 
restricted to be one radius away from the boundaries of the lattice,
so that full bubbles are nucleated, bubble radius being about 13.8.
In contrast, in \cite{ajt}, nucleation sites were restricted to be 5 
bubble radii away from the lattice boundaries to avoid spurious reflections, 
due to use of free boundary conditions there.] As we will see below, the 
separation of vortices and antivortices in these pairs is now much larger, 
and these defects last for much longer times. 

 We now give some specific examples of vortex-antivortex pairs 
formed due to flipping of $\Phi$. Fig.2 shows the plot of $\Phi$ for
the entire lattice somewhat after the onset of phase transition. 
The bubbles have grown in size and some of them have collided.
Fig.3a shows the $\Phi$ plot at t = 49.8 of a Kibble mechanism vortex 
located at x = 62.8, y = 151.6. The vortex is somewhat distorted and
was formed by the collapse of a region of false vacuum, surrounded by true
vacuum, with a net winding trapped in it. Such situations were also observed
in \cite{ajt} where the trapped false vacuum region was seen to assume
spherical shape as it gradually collapsed. Due to the nature of formation 
of the vortex, the field near its core  oscillates resulting in the flipping 
of $\Phi$ and subsequent formation of a vortex-antivortex pair due to 
flipping; as shown in Fig.3b at t = 52.0. The Kibble vortex is the right
most one, at x = 65.8, y = 151.0, while the vortex and the antivortex created 
due to flipping are towards left of it, at x = 58.6, y = 153.4 and x = 62.2, 
y = 149.8, respectively. Note the flipped orientation of $\Phi$ in between 
this vortex and antivortex, compared to the orientation of $\Phi$ in the same 
region in Fig.3a. The vortex and the antivortex in this pair are well
separated, by a distance equal to about 5, which is roughly twice the 
string core radius. Fig.3c shows the plot of $\Phi$  of the same region at 
t = 54.3. The anti-vortex belonging to the pair has moved closer to the 
Kibble mechanism vortex; (eventually annihilating it and leaving
behind the {\it vortex belonging to the pair}).

 There is another instance in which oscillations of the core of an
antivortex formed via the Kibble mechanism give rise to a vortex-antivortex 
pair by the flipping mechanism. Here also, the Kibble antivortex forms
due to the collapse of a region of false vacuum having net antiwinding.
In this case, we find that the Kibble mechanism anti-vortex
subsequently annihilates with the vortex belonging to the pair thereby 
leaving behind the {\it anti-vortex of the pair}. We consider the 
vortex/anti-vortex which survives to be effective Kibble mechanism 
vortex/antivortex and count it as such. These examples clearly show that
the vortex and antivortex formed by this mechanism can be separated far
enough to mix with the Kibble mechanism vortices and hence affect the 
defect distribution formed via the Kibble mechanism.

 In Figs.4a-4b we give another example of vortex-antivortex pairs forming 
due to flipping. Fig.4a gives the $\Phi$  plot in a region in which there 
is no net windingi, though there are strong phase gradients. The plot of 
$\Phi$ of the same region at 
t = 72.4 shows a vortex-antivortex pair, Fig.4b. Comparison of the two 
plots clearly indicate that the vortex-antivortex pair is due to flipping.
This is also confirmed by the surface plots of $\phi$. The vortex-antivortex 
pair is well formed and reasonably well separated.  

 Figs.5a-5f show a series of plots depicting the
formation (by the flipping mechanism) and evolution of two well separated 
vortex-antivortex pairs as well as a cluster of overlapping vortex-antivortex 
pairs. The vortex-antivortex pairs belonging to the cluster annihilate soon 
after formation. Fig.5a shows the $\Phi$ plot of the region, at t = 67.9, 
in which there is no net winding or antiwinding present but there are huge 
field oscillations (as confirmed by the surface plots). A triangular region 
of false vacuum is seen near the top left region together with two other 
oscillating regions. Fig.5b shows the plot of $\Phi$ of the same region at 
t = 74.6. Comparison with Fig.5a clearly shows that the thin strip of region
seen in Fig.5a has flipped resulting in the formation of a vortex-antivortex    
pair. The vortex is at x = 188.8, y = 156.7, and has a well formed core while 
the antivortex is below it, towards left, and has a very large
core where field is oscillating. The triangular region of false vacuum seen 
in Fig.5a has almost disappeared by now. [Oscillations brought about by 
collision of bubble walls leads to the formation of yet another 
vortex-antivortex pair due to flipping at t = 70.1. However, this pair is 
highly overlapping and annihilates soon. We therefore do not count it
as a pair, since we are only interested in formation of those defects
which can survive at least for a while, in order that they can affect
the defect distribution. As we mentioned earlier, our criterion for counting 
a given vortex- antivortex pair is that the separation should be larger than 
the core thickness of the string.]  

Fig.5c shows the same region at t = 79.2. Apart from the original pair, 
another pair is seen just below the vortex of the original pair, with the
vortex and the antivortex of this second pair being separated along the x 
axis, with y $\simeq$ 153.8. The core of anti-vortex, belonging to the 
original pair, has shrunk, though the anti-vortex is still distorted. 
Also, the formation of the new pair has led to further separation of the 
vortex and the antivortex belonging to the original pair, the separation 
being about 23 units at this stage. This is a very large separation, being
almost an order of magnitude larger than the string core radius.
Fig.5d shows the plot of $\Phi$ at t = 81.5. Both the pairs are now 
well-formed and well separated.

 Subsequent $\Phi$ plot at t = 86.0 (Fig.5e) shows the vortex belonging 
to the newly formed pair moving to the anti-vortex belonging to the 
original pair (and ultimately annihilating). The field oscillations generated 
by the annihilation of the pair leads to two new vortex-antivortex pairs
(due to flipping) in the region. This is observed in the $\Phi$ plot of
this region at t = 90.5. However, we do  not count these pairs as they are 
not well separated and annihilate very soon. Fig.5f shows the plot of $\Phi$ 
of this region at t = 90.5, showing an isolated vortex and a region with
net antiwinding. The vortex belonging to the original pair as well as the
anti-vortex belonging to the pair formed at t = 79.2 do not annihilate till 
the end of the simulation, i.e. till t = 101.8. Of all the pairs formed
due to flipping, this is the one which survives the longest.
   
\vskip .3in
\centerline{\bf 5. Discussion of the Results}
\vskip .1in

 We have mentioned earlier that in order to have pair-production due to
flipping of $\Phi$, the bubble collisions must be sufficiently energetic.
This is only possible for low nucleation rates for which bubble walls can
acquire sufficient kinetic energy, before collision, by the conversion of 
false vacuum to true vacuum. Results obtained in \cite{dss} (the zero
explicit symmetry breaking case), for a specific initial configuration, 
seemed to indicate that a pair production, where vortex and antivortex 
are well separated, may not occur too often. However, our results in this 
paper clearly indicate that even in a realistic phase transition where 
bubbles nucleate randomly, this new mechanism may become very prominent
and may be the deciding factor in determining the defect distribution.
In this context we mention that we find that the dynamics of collisions
of several bubbles often conspires to enhance the magnitude of field 
oscillations thereby making flipping of $\Phi$ easier.

 Another intriguing feature observed in our simulations is the presence of
vortices/antivortices with an oscillating core. This feature is especially
prominent in vortices/antivortices formed by the collapse of a large
region of false vacuum  in which a winding/anti-winding is trapped.

   It is important to note that the results in \cite{dgl} had shown
that for systems with explicit symmetry breaking, this mechanism
can produce very large number of defects. However, such systems are
very special and in most cases (in condensed matter systems, or in
particle physics, especially for models of structure formation in the 
early Universe) one is interested in defect production in systems {\it
without} any explicit symmetry breaking. From this point of view, our
present results are important as they demonstrate that for very low 
nucleation rates this mechanism may be most dominant for defect production.
For example, this mechanism may completely dominate if one is interested 
in studying defect production at last stages of  extended inflation in 
the early Universe \cite{stein}  where bubble expansion is not impeded 
by damping. Cosmic string formation by Kibble mechanism has been studied 
in extended inflationary models in \cite{kolb}. There it was argued that 
the correlation length, taken to be the mean bubble size at the end of 
inflation, is larger than that corresponding to the Kibble mechanism for
a thermal second order transition, and that this would result in the 
formation of a more dilute network of strings. In view of our results in 
this paper, one expects to see a fairly large number of small string
loops in addition to the Kibble mechanism cosmic strings. [As we have 
mentioned, in three dimensions, the flipping of $\Phi$ will result in 
the formation of string loops \cite{dss}.] This can drastically 
alter the number density of defects and may lead to a much denser
network of strings.
       
 For high nucleation rates and in the presence of damping, this mechanism 
will be considerably suppressed. This follows because the bubble walls will
acquire less energy before collision thereby reducing the magnitude of $\Phi$
oscillations and the probability of flipping of $\Phi$. In simulations with
high nucleation rates, we find some highly overlapping pairs (as were
found in \cite{ajt}). Presence of these indicates 
that the magnitude of field oscillations is large enough
to induce $\phi$ to pass through $\phi=0$ thereby resulting in flipping, but
{\it not} large enough to take $\phi$ all the way across the barrier to the 
other side of the effective potential which would result in the formation 
of a well separated pair. 

The vortex-antivortex pairs eventually annihilate because of the attractive 
force between them. Even then, as we have shown above, 
in certain cases, density waves generated by field oscillations
from neighboring regions, as well as the presence of other vortices lead 
to the separation of the pair, thereby delaying their eventual annihilation
for a significant period of time. For three dimensions, this will imply 
formation of expanding string loops \cite{dss}. A dense network of such 
string loops can lead to formation of very long strings via intercommuting 
of entangled loops. In any case, a dense network of string loops will
certainly modify the network of strings and hence can affect its
subsequent evolution. 

\vskip .3in
\centerline {\bf 6. Conclusions}
\vskip .1in
 
We have carried out numerical simulations of a general first order phase 
transition for the case of spontaneous breaking of a global U(1) symmetry,
and have studied the production of vortices and antivortices.  We estimate 
the net number of vortices produced, which includes vortices formed due to 
the Kibble mechanism as well as those produced via the pair production 
mechanism.  Nucleation rate affects density of defects produced via the
flipping mechanism due to the fact that a larger nucleation rate implies 
smaller average bubble separation, which in turn leads to less kinetic 
energy for the bubble walls before bubbles collide. Oscillations of 
$\phi$ are less prominent for less energetic walls leading to smaller 
number of defect-antidefect pairs for larger nucleation rates. 

  We therefore simulate the transition with a low nucleation rate.
Here the bubble collisions are energetic enough to lead to large
oscillations of $\phi$ and subsequent flipping of $\Phi$. This leads to 
the production of many, well separated, vortex-antivortex pairs
via the new mechanism. For example, we find a total of 14, reasonably
well separated, vortices and antivortices formed via this mechanism, 
as compared to 6 vortices and antivortices formed due to the Kibble 
mechanism. These results demonstrate that for very low nucleation
rates, when bubble collisions are extremely energetic, this mechanism
may drastically alter the defect production scenario. A dense
network of defects produced via this mechanism can completely
modify the network of strings produced via the Kibble mechanism
and hence may alter the evolution of string network. This may be the 
situation for inflationary theories of the early Universe, for example,
in extended inflation, where bubble collisions 
are very energetic. In fact in view of our results, one can
expect a large population of other defects, especially monopoles,
arising via this mechanism at the end of extended inflation.

 Interestingly, a first order transition with low nucleation rate would 
imply large bubble separations and hence a smaller number of Kibble
defects (say monopoles). However, defects (per bubble) produced via field 
oscillations are more abundant for low nucleation rate due to collisions 
being more energetic. Therefore, the final defect density may well be an
increasing function of the bubble separation (say in extended inflation).
In that case there seems a possibility of overproducing monopoles. This
interesting possibility needs further exploration.
 
 A direct experimental evidence for this mechanism can only come from 
condensed matter systems, as was the case for the Kibble mechanism
\cite{zurk}. The phase transition in superfluid $^3$He from A to B phase
is of first order, and occurs via nucleation of bubbles of true vacuum, 
the growth of which is unimpeded by damping. Hence, this mechanism should 
lead to formation of small string loops in this transition. In view of our 
results in this paper, we expect that number density of such string loops 
may be significant. It will be very exciting to detect these loops. As we 
had emphasized in \cite{dss}, observation of loops smaller than the average 
size of coalescing bubbles, at the string formation stage, will give direct 
evidence for this mechanism.
         

\begin{figure}[h]
\begin{center}
\vskip -1.5 in
\leavevmode
\epsfysize=12truecm \vbox{\epsfbox{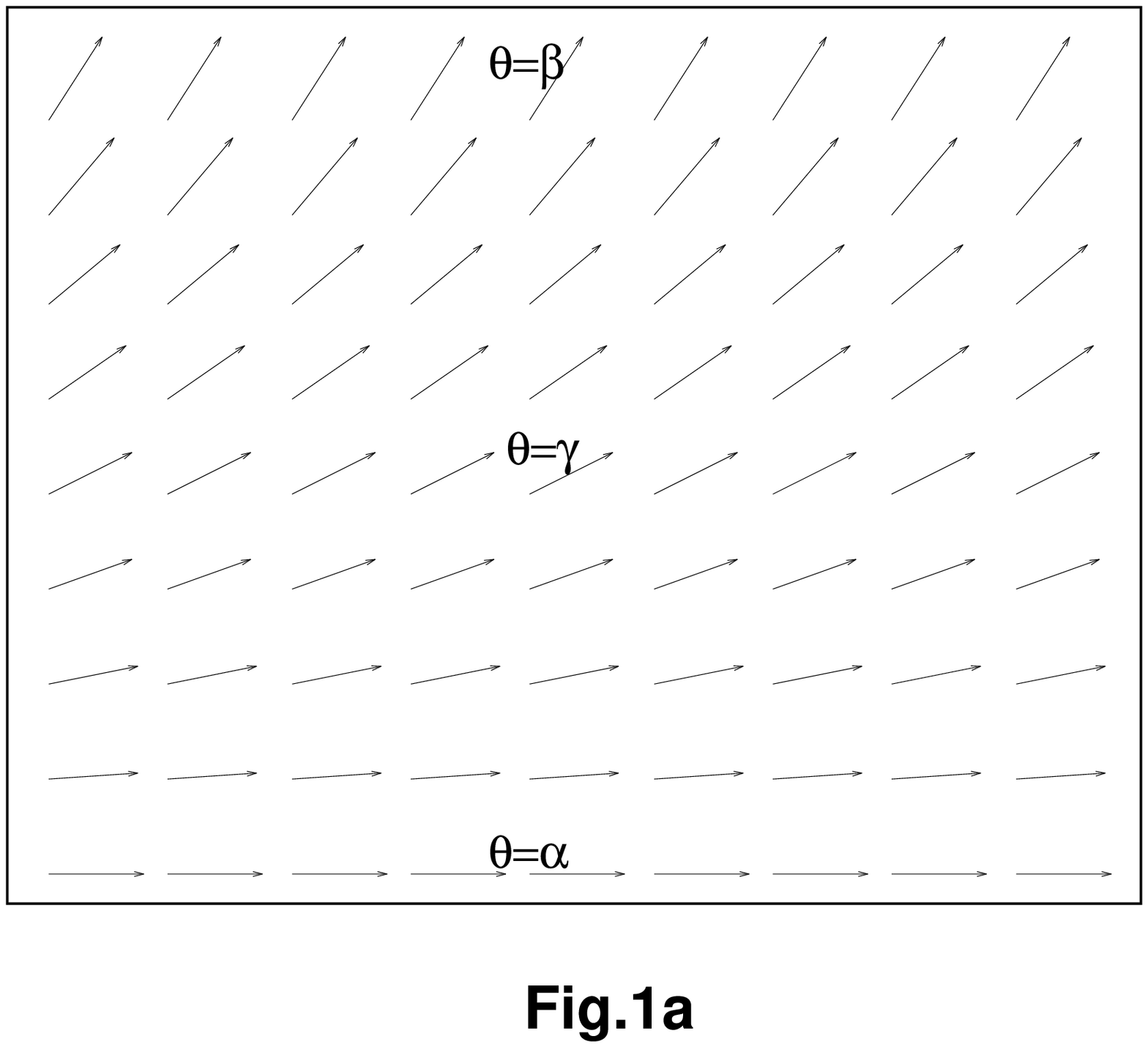}}
\vskip -1.5in
\end{center}
\end{figure}
\begin{figure}[h]
\begin{center}
\vskip -1.5 in
\leavevmode
\epsfysize=12truecm \vbox{\epsfbox{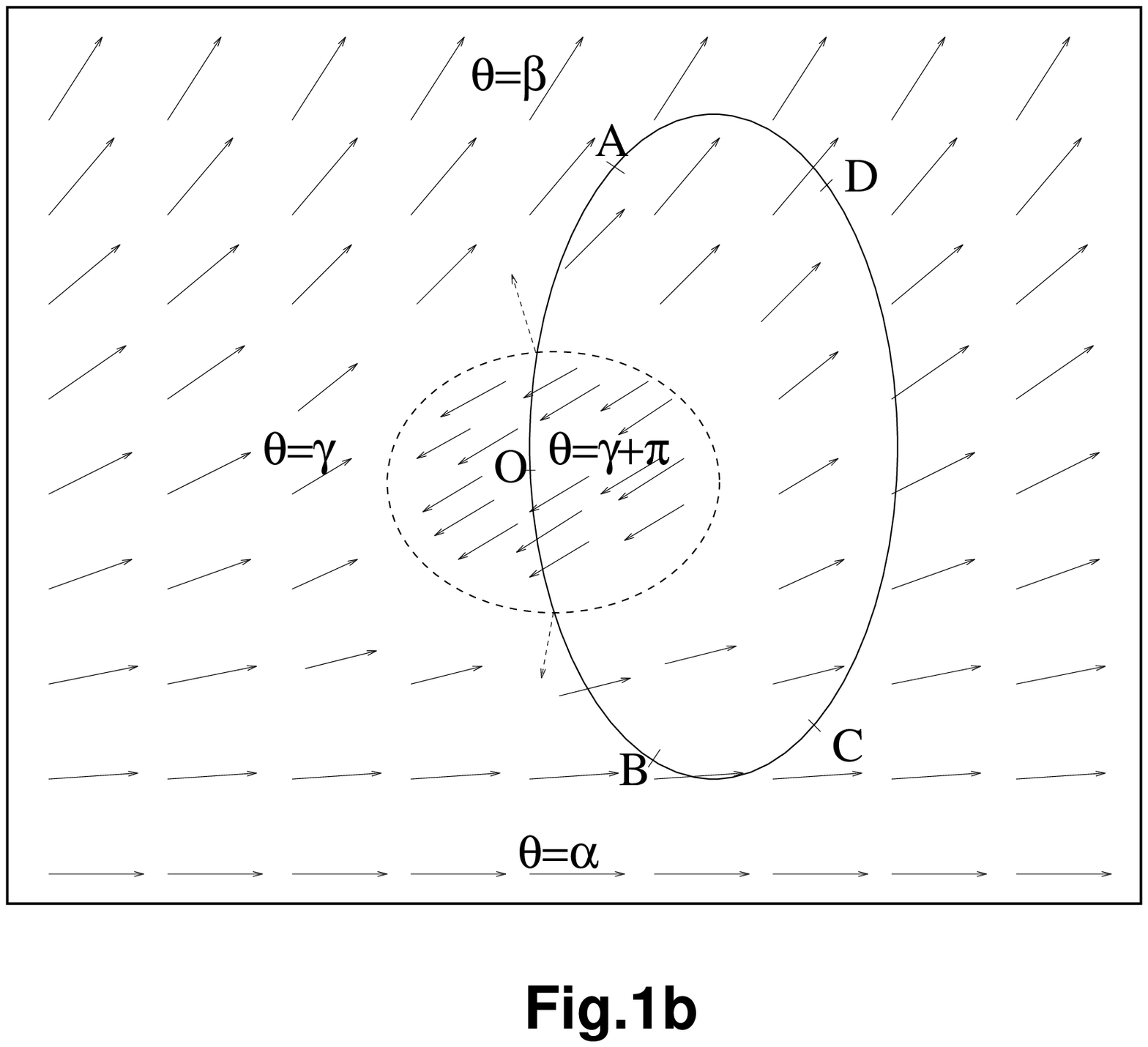}}
\vskip -1.5in
\end{center}
\caption{(a) A region of space with $\theta$ varying uniformly
from $\alpha$ at the bottom to some value $\beta$ at the top.
(b) Flipping of $\Phi$ in the center (enclosed by the dotted
loop) has changed $\theta = \gamma$ to $\theta = \gamma + \pi$ 
resulting in a pair production.}
\label{Fig.1} 
\end{figure}

\begin{figure}[h]
\begin{center}
\vskip -0.5 in
\leavevmode
\epsfysize=12truecm \vbox{\epsfbox{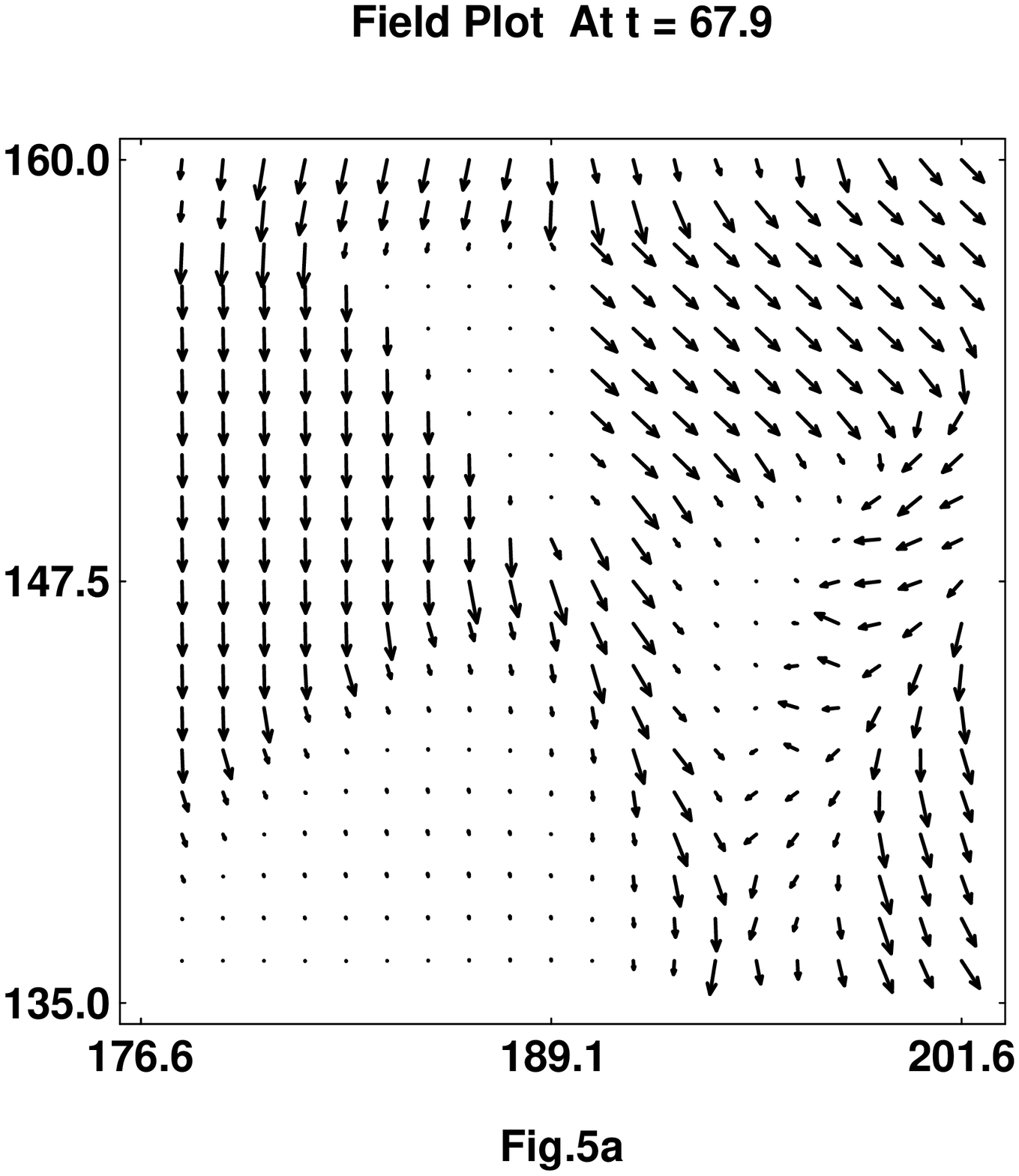}}
\vskip -0.5in
\end{center}
\end{figure}

\begin{figure}[h]
\begin{center}
\vskip -0.5 in
\leavevmode
\epsfysize=12truecm \vbox{\epsfbox{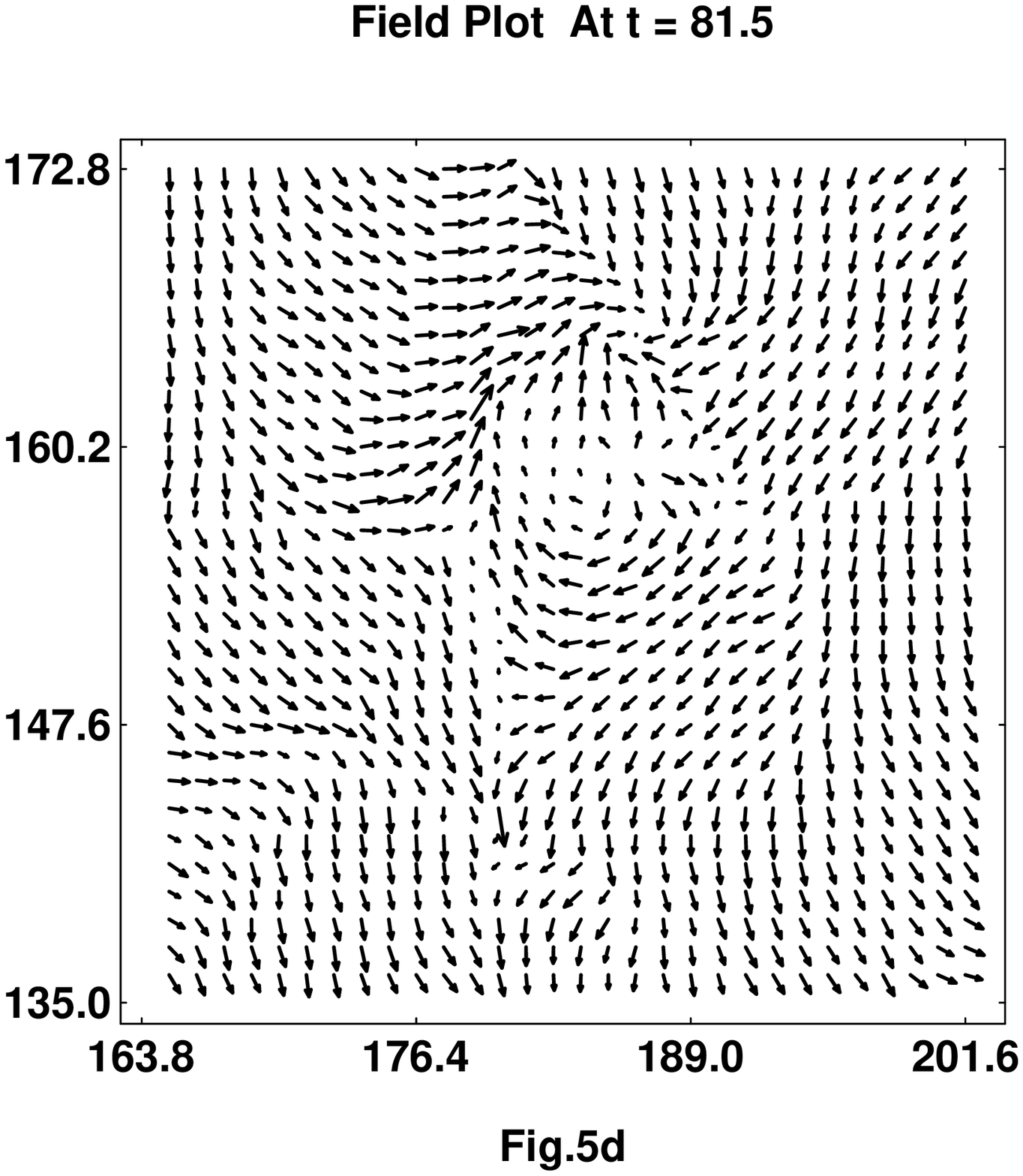}}
\vskip -0.5in
\end{center}
\end{figure}

\end{multicols} 
\end{document}